\documentclass[
aps,
12pt,%
final,%
notitlepage,%
oneside,%
twocolumn,%
superscriptaddress,%
noshowpacs,%
centertags]%
{revtex4}
\usepackage{graphicx}

\newcommand{\aver}[1]{\left\langle#1\right\rangle}
\newcommand{\cat}[1]{\left|#1\right\rangle}
\newcommand{\brc}[2]{\left\langle#1\left|#2\right|#1\right\rangle}
\newcommand{\bc}[2]{\left\langle#1\left.\right|#2\right\rangle}

\begin{document}
\selectlanguage{english} 
\title{Imaginary Entanglement as Cost of Unitarity}
\author{Constantin V. Usenko}
\email[]{usenko@univ.kiev.ua}
\affiliation{National Shevchenko University of Kyiv, Theoretical Physics Department}
\begin{abstract}
This report is about contradiction between fidelity  needed to determine the entanglement and concomitant noise that always accompanies precise measurement.

Account of quantum properties of field leads to additional noise caused by multiple particle creation  through nonunitarity of quantum field representation in embedded sections of space (Unruh noise).

Causes of quantum noise vanish at leaving off assumption about statistical independence of detectors. Smearing of detector leads to elimination of causes of Unruh noise and to emergence of imaginary entanglement of few mode states caused by overlap of detector sections.
\end{abstract}
\maketitle
\section{Introduction}
Progress in creation of essentially quantum states of electromagnetic field (entangled states of few photons) leads to need in analysis of physical meaning of the basic concept of quantum physics - concept of probability distribution of particle registration \cite{amp,bjss} by several placed near each other detectors.

Well known mathematical scheme of determination of probability distribution deals with probabilities of detection of particles in the sequence of embedded sections (Rokhlin scheme) in implicit assumption that detectors are statistically independent. One expects that at decrease of the size of detector count decreases proportionally. The constant of proportionality is taken as probability density (probability per size).

Account of quantum properties of field leads to additional noise \cite{noise} caused by multiple particle creation \cite{pvmc,bjfp} through nonunitarity of quantum field representation in embedded sections of space \cite{vacent} (Unruh noise).

Causes of quantum noise vanish at leaving off assumption about statistical independence of detectors. Smearing of detector leads to elimination of causes of Unruh noise and to emergence of imaginary entanglement of few mode states caused by overlap of detector sections.

This report is about contradiction between fidelity  needed to determine the entanglement and concomitant noise that always accompanies precise measurement.

Specific objects in the study of entanglement of states are pairs of polarized photons (bi-photons especially) and electrons with opposite spins.

Typical procedure of study of entanglement is in consideration of entangled electron pair moving to pair of spin state detectors. There exists some probability for the state "UP-Down" and some probability for the opposite state.

Identity of particles is one more reason for entanglement.
The process in which the upper particle is registered by down detector and
vice versa is identical to process in which each particle falls to its
own detector.
\section{Entanglement in Measurement for Two Parts}
Mathematical model is based on the study of the density matrix properties.

Mathematical model of the measuring unit is expansion of unity forming positive operator-valued measure in the space of states of the system.
The device measuring the states of the first sub-system does not detect the state difference of the second one, and vice versa.

\[
    \begin{array}{c|c}
 M_A = M^{(a)} \otimes 1^{(b)}& M_B = 1^{(a)} \otimes M^{(b)}\\
 M^{(a)}\left[ {\hat {\rho }_1^{(a)} } \right] = M_{a1}
 & M^{(b)}\left[ {\hat {\rho }_1^{(b)} } \right] = M_{b1}
  \\
 M^{(a)}\left[ {\hat {\rho }_2^{(a)} } \right] = M_{a2}
 &M^{(b)}\left[ {\hat {\rho }_2^{(b)} } \right] = M_{b2} \\
 \end{array}
\]
\subsection{Independent parts}
The system consisting of independent parts has as density matrix direct product of density matrices of separate parts
\[
\begin{array}{ll}
    \hat {\rho }^{(a + b)} =& 
     \left( {p_a \hat {\rho }_1^{(a)} + \left( {1 - p_a
    } \right)\hat {\rho }_2^{(a)} } \right) \\
    &\otimes \left( {p_b \hat {\rho
    }_1^{(b)} + \left( {1 - p_b } \right)\hat {\rho }_2^{(b)} } \right).
\end{array}
\]
As the result, reaction of the device on the system consisting of independent parts is equal to the reaction on its own sub-system. Probability of registration of some state of the system is equal to the product of probabilities for respective states of the sub-systems
$P\left( {a1} \right) = p_a $, $ P\left( {a2} \right) = \left( {1 - p_a }
\right)$, $
 P\left( {b1} \right) = p_b $, $P\left( {b2} \right) = \left( {1 - p_b }
\right)$.
\[
    \begin{array}{l}
 P\left( {a1\& b1} \right) = p_a p_b ;\\ 
 P\left( {a2\& b1} \right) = \left( {1 - p_a } \right)p_b; \\
 P\left( {a1\& b2} \right) = p_a \left( {1 - p_b } \right);\\ 
 P\left(a2\& b2\right) = \left(1 -p_a\right)\left(1 - p_b\right);  \\
P\left( {a_k\& b_m} \right) =P\left( {a_k} \right)P\left( {b_m}\right).
 \end{array}
\]
\subsection{Entangled state}
Entangled states are characterized by density matrix being unreducible
mixture of the products of sub-system states
\begin{equation}
    \hat {\rho }^{(a + b)} = p\hat {\rho }_{1a}^{(a)} \otimes \hat {\rho
}_{1b}^{(b)} + \left( {1 - p} \right)\hat {\rho }_{2a}^{(a)} \otimes \hat
{\rho }_{2b}^{(b)}
\end{equation}

Probability of registration of some state of one sub-system essentially
depends on the result of measurement for another one, and joint probability
distribution is different from product of probability distributions for each
of the sub-systems
\[
\begin{array}{l}
 P\left( {a1\& b1} \right) = p \ne P\left( {a1} \right)P\left( {b1}
\right);\\
 P\left( {a2\& b1} \right) = 0 \ne P\left( {a2} \right)P\left(
{b1} \right); \\
 P\left( {a1\& b2} \right) = 0 \ne P\left( {a1} \right)P\left( {b2}
\right);\\
 P\left( {a2\& b2} \right) = \left( {1 - p} \right) \ne P\left(
{a2} \right)P\left( {b2} \right),
 \end{array}
 \]
and mutual probapilities are
\[
    \begin{array}{cc}
     P\left( {a1\vert b1} \right) = 1;& P\left( {a2\vert b1} \right) = 0; \\
     P\left( {a1\vert b2} \right) = 0;& P\left( {a2\vert b2} \right) = 1.
     \end{array}
\]

\section{State Discrimination}
Detection area is divided between counters.
It is supposed that each counter registers one of the possible states of the
system, and there exists one-to-one correspondence between the counters and
the states.

Deviations from such correspondence are taken as noise.
\subsection{Event space bisection}

Definition of probability density is constructed through sequence of
bisections of space of events (Rokhlin scheme).

Probability distribution on the event space, according to axiomatics of
probability theory, is realized through proceeding to limit in the sequence
of bisections of the event space. The initial element of the sequence is the
reliability partition -- the subset of the event space for which probability
of any outside event is equal to zero. It is supposed that such area has
finite measure.

Bisection of the event space is accompanied by adjustment of the methods for
description of the states of the particles being detected.

Each section of event space has its own set of states localized in that
section.

Now we consider common coordinate space as event space. Description of
particles in each section is performed here by means of a set of wave
functions localized in respective section.

\subsection{Quantum field on section}

\begin{figure}[ht]
\setcaptionmargin{15mm}
\onelinecaptionsfalse 
\includegraphics[width=1.8in]{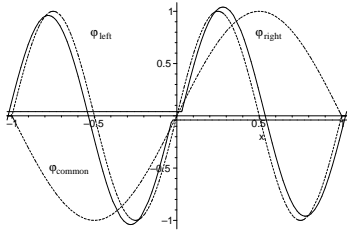}
\captionstyle{normal}\caption{ Wave functions on a section and its subsections.
Wave function of whole section $\phi_{common}$ and the same of subsections - $\phi_{left}$ which vanishes on right subsection and $\phi_{right}$ - vanishes on left one are shown. }
\label{wf}
\onelinecaptionsfalse 
\end{figure}

In the graph (\ref{wf}) typical section of length $2L$ is shown, and expression for the spatial mode with quantum number $m$ is
$\phi _m \left(x,t\right) = \frac{1}{\sqrt L }e^{\left( ip_m x -
i\varepsilon \left( p_m  \right)t \right)}$ where $p_m = \frac{\pi m}{L}$ and $\varepsilon \left( p \right) \equiv \varepsilon _p = \sqrt {\mu ^2 + p^2}$.
Quantum field following Fermi statistics is convenient because of
limited number of particles in each mode.

\begin{equation}
\begin{array}{l}
	\hat{\psi}\left(x,t\right)
	=\\
	\sum_{m=-\infty}^\infty{
	\left(
	 u_{p_m}^{(+)} \phi_m\left(x,t\right)\hat{a}_m + u_{p_m}^{(-)} \phi _m^\ast \left(x,t\right)\hat{b}_m^+
	\right)
	}
	\end{array}
	\label{df}
\end{equation}

The spinors
 \[u_p^{(+)}
 =\left(\begin{array}{cc} \frac{\varepsilon + \mu }
 {\sqrt {2\varepsilon \left(\varepsilon + \mu \right) }}  \\ \frac{p}{\sqrt {2\varepsilon \left(\varepsilon + \mu \right)}} \end{array}
 \right); \quad
u_p^{(-)} = \left(\begin{array}{c}
\frac{ - p}{\sqrt {2\varepsilon \left(\varepsilon + \mu \right) }}\\
\frac{\varepsilon + \mu }{\sqrt {2\varepsilon \left(\varepsilon +
\mu\right) }}
\end{array}\right),\]
describe separation of states to positive-frequency and
negative-frequency ones.

Hamiltonian of the field is

\[
	\hat{H}=\sum_{m =-\infty}^\infty {\hbar \varepsilon \left( p_m\right)
	\left(\hat{a}_m^+ \hat{c} _m+\hat{d} _m^+ \hat{b}_m \right)}
\]
Operators $\hat{a}_m$, $\hat{a}_m^+$ generate Heisenberg-Weil algebra for particle modes and $\hat{b}_m$, $\hat{b}_m^+$ for antiparticle ones.
\subsection{Quantum field on bisected bases}

Bisection of the initial section of length 2L into two parts requires
construction of additional representation of quantum field in each of the
subsections. Wave functions for the left subsection are shown in the graph (\ref{wf}).

Decrease of length leads to re-definition of the sequences of momenta for each of the modes. The new sequence is $\left\{q_m=\frac{2\pi m}{L}:\quad m \in Z\right\}$.

Representation in left subsection is given by expansion of quantum field
\begin{equation}
\begin{array}{ll}
	\hat{\psi}\left(x,t\right)
	=&
	\sum_{m=-\infty}^\infty{
	\left(
	 u_{q_m}^{(+)} \phi_{m,left}\left(x,t\right)\hat{c}_m \right.}\\
	& {\left.+ u_{q_m}^{(-)} \phi _{m,left}^\ast \left(x,t\right)\hat{d}_m^+
	\right)
	}
	\end{array}
\end{equation}
Heisenberg-Weil algebra for that representation is generated by operators $\hat{c}_m$, $\hat{c}_m^+$ and $\hat{d}_m$, $\hat{d}_m^+$.
\[
\hat{H}
=\sum_{m =-\infty}^\infty {\hbar \varepsilon \left( q_m\right)
\left(\hat{c}_m^+ \hat{c} _m+\hat{d} _m^+ \hat{d}_m \right)}
\]

Bisection leads to one more representation associated to the right subsection. 
\begin{equation}
\begin{array}{ll}
	\hat{\psi}\left(x,t\right)&
	=\sum_{m=-\infty}^\infty{
	\left(
	 u_{q_m}^{(+)} \phi_{m,right}\left(x,t\right)\hat{f}_m  \right.} \\
	 &{\left.+ u_{q_m}^{(-)} \phi_{m,right}^\ast \left(x,t\right)\hat{g}_m^+
	\right)
	}
	\end{array}
\end{equation}
Heisenberg-Weil algebra for that representation is generated by third set of operators $\hat{f}_m$, $\hat{f}_m^+$ and $\hat{g}_m$, $\hat{g}_m^+$.
\[
\hat{H}
=\sum_{m =-\infty}^\infty {\hbar \varepsilon \left( q_m\right)
\left(\hat{f}_m^+ \hat{f} _m+\hat{g} _m^+ \hat{g}_m \right)}
\]

\subsection{Bogolubov transform}

The operators of creation and annihilation have changed along with 
perception of particles and antiparticles.

Inverse Fourier transform for subsection bases 
\[\begin{array}{c}
\hat{c}_m=\bc{u_{p_m}^{(+)}\phi_{m,left}\left(x,t\right)} {\hat{\psi}\left(x,t\right)};\\
\hat{d}_m=\bc{u_{p_m}^{(-)}\phi_{m,left}\left(x,t\right)} {\hat{\psi}^+\left(x,t\right)}\\
\hat{f}_m=\bc{u_{p_m}^{(+)}\phi_{m,right}\left(x,t\right)} {\hat{\psi}\left(x,t\right)};\\
\hat{g}_m=\bc{u_{p_m}^{(-)}\phi_{m,right}\left(x,t\right)} {\hat{\psi}^+\left(x,t\right)}
\end{array}
\]
and direct Fourier transform for section basis (\ref{df}) produce interdependence between operators of creation and annihilation
\begin{equation}
\begin{array}{c}
	\hat{c}_m 
= \sum_{n = - \infty }^\infty {
\left(\alpha _{m,n}\hat{a}_n + \beta _{m,n}^\ast\hat{b}_n^+ \right)
};\\ 
 \hat{d}_m 
= \sum_{n = - \infty }^\infty {\left( -\beta_{m,n}^\ast 
\hat{a} _n^+ + \alpha_{m,n}\hat{b} _n\right)}; \\
	\hat{f}_m 
= \sum_{n = - \infty }^\infty {
\left(\alpha _{m,n}\hat{a}_n + \beta _{m,n}^\ast\hat{b}_n^+ \right)
};\\ 
 \hat{g}_m 
= \sum_{n = - \infty }^\infty {\left( -\beta_{m,n}^\ast 
\hat{a} _n^+ + \alpha_{m,n}\hat{b} _n\right)} 
\end{array}
\end{equation}

One of good effects of bisection is in almost complete coincidence of the Bogolubov transform coefficients for the left and right subsections
\[\begin{array}{l}
\alpha _{m,k} = \frac{1}{\sqrt{ 2 }}\delta _{2m,k}
 - \frac{i}{\sqrt{ 2 \pi }}\sum_{n = - \infty }^\infty {
A_{n,m,k} \delta _{2n+1,k}
  } ;
\\
\beta _{m,k} = W_m\delta _{ - 2m,k}
 - \frac{i }
{\sqrt {2 \pi }}\sum_{n = - \infty }^\infty {
B_{n,m,k}\delta _{2n+1,k} } ;\\
A_{n,m,k}=\\\frac{\left(\varepsilon \left(p_k\right) + \mu\right) \left(\varepsilon \left(q_m\right) + \mu\right) + p_k q_m }
 {2\sqrt {
 \varepsilon \left(p_k\right)\varepsilon \left(q_m \right)
 \left(\varepsilon \left(p_k\right) + \mu\right)
 \left(\varepsilon \left(q_m\right) + \mu\right)} }
 \frac{e^{i\left(
 \varepsilon\left(q_m\right) - 
 \varepsilon\left(p_k\right)\right)t}}{n - m + 1/2};\\
B_{n,m,k}=\\
\frac{p_k \left(\varepsilon \left(q_m\right) + \mu \right) 
- q_m \left(\varepsilon\left(p_k\right) + \mu\right)}
{2\sqrt {\varepsilon \left(p_k\right)\varepsilon \left(q_m\right)
\left(\varepsilon\left(p_k\right) + \mu\right)
\left(\varepsilon\left(q_m\right) + \mu\right)} }
\frac{e^{ - i\left(\varepsilon\left(q_m\right) + 
\varepsilon \left(p_k\right)\right)t}}
{n + m +1/2};\\
W_m=\frac{q_m e^{ - 2i\varepsilon \left( {q_m } \right)t}}{\sqrt 
2 \varepsilon \left( {q_m } \right)}.
\end{array}
\]

\section{Fermi Particle Creation}
Vacuum state $\cat{0}$ for section basis $\hat{a}_m\cat{0}=0$, $\hat{b}_m\cat{0}=0$
can not be vacuum for subsection basis since $\hat{c}_m\cat{0}=\sum{\beta_{m,n}b^+_n\cat{0}}\neq 0$.

Entanglement of the modes of particles and antiparticles leads to unitary nonequivalence of the representations under consideration.
The state being vacuum with respect to all the annihilation operators of the section is not vacuum with respect to the operators of the left subsection. As the result the vacuum average of the particle or antiparticle number of each mode of the subsection is non-zero.
\begin{equation}
\begin{array}{r}
	\aver{\hat{n}_k}
	=\brc{0}{\hat{c}^+_k\hat{c}_k}=\brc{0}{\hat{d}^+_k\hat{d}_k}\\
	=\brc{0}{\hat{f}^+_k\hat{f}_k}=\brc{0}{\hat{g}^+_k\hat{g}_k}\\
	=\sum_{n=-\infty}^{\infty}{\left|\beta_{n,k}\right|^2}
	\end{array}
\end{equation}

\begin{figure}[ht]
\setcaptionmargin{15mm}
\onelinecaptionsfalse 
\includegraphics[width=1.8in]{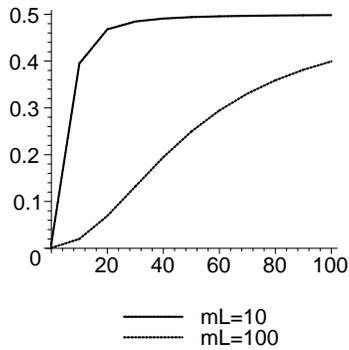}
\captionstyle{normal}\caption{Particle creation as bisection result. Average value of particle number versus mode number. Parameter $mL$ is equal to the ratio of the linear dimension of the section to the Compton wavelength. }
\label{creation}
\end{figure}

The graph (\ref{creation}) shows the dependence of the filling numbers on the mode number. Specific parameter is here the ratio of the linear dimension of the section to the Compton wavelength. With decrease of this ratio the graph goes up near origin more abruptly and more quickly comes to saturation value 0.5.
Saturation value corresponds to our notion about total stochasticity of the phenomenon.
\subsection{ Correlation of Noise}
Since the vacuum state of initial section is a pure state, stochasticity of mode filling in subsection representations makes evidence of correlation between noises.
The most interesting correlation is correlation between filling numbers for the modes of the left and the right subsections.
\begin{equation}
\begin{array}{r}
	D_{c,k;f,m} =\brc{0}{\hat{c}^+_k\hat{c}_k\hat{f}^+_m\hat{f}_m}\\
	-\brc{0}{\hat{c}^+_k\hat{c}_k}\brc{0}{\hat{f}^+_m\hat{f}_m}\\
	=\sum_{l=-\infty}^{\infty}{\beta_{l,k}\beta^*_{l,m}}
	\sum_{n=-\infty}^{\infty}{\alpha_{n,m}\alpha^*_{n,k}}
	\end{array}
\end{equation}

\begin{figure}[ht]
\setcaptionmargin{15mm}
\onelinecaptionsfalse 
\includegraphics[width=1.8in]{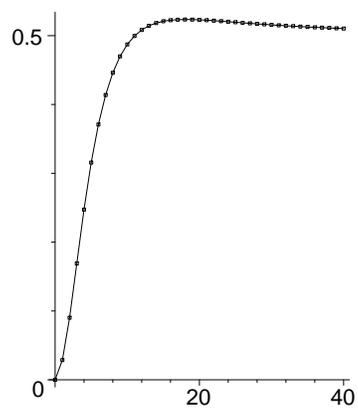}
\captionstyle{normal}\caption{Correlation function between filling numbers for the modes of the left and the right subsections.}
\label{cor}
\end{figure}

 Correlation function is substantially non-zero for coinciding mode numbers only $D_{c,m;f,n} \approx \delta _{m,n} D_n $ and comes to values corresponding to almost total correlation between the modes of the subsections shown by graph(\ref{cor}).
 
So, the states of the particles in the left and the right subsections are entangled. 

Unitary non-equivalence of representations in subsections results from restriction of wave packets.
\subsection{Overlap of bases }
Alternative version is in field description by means of sets of states with incomplete localization.
Each such state has a respective wave packet with well-defined average values of coordinate and momentum, and each set of states with equal average values gives quantum field representation associated with a given point of phase space, representations are distributed over the whole phase space.
As such a set we use a set of oscillator states
\begin{equation}
	\begin{array}{ll}
	\varphi_\alpha\left(p\right)&
	=\frac{1}{\sqrt{\sigma\sqrt{2\pi}}}
	\exp{\left(-\frac{\left(p-p_\alpha\right)^2}{4\sigma^2}-ipx_\alpha\right)}\\
	\varphi_{\alpha,n}\left(p\right)&
	=\frac{\left(\sigma\frac{d}{dp}+\frac{p}{2\sigma}-\sigma x_{\alpha}
	+i\frac{p_{\alpha}}{2\sigma}\right)^n
	}{\sqrt{n!}}\varphi_\alpha\left(p\right)
	\end{array}
\end{equation}
Here $\alpha=\sigma x_{\alpha}	+ip_{\alpha}/2\sigma$, and overlap of basis functions is given by 
\[
\begin{array}{l}
\int{\varphi^*_\beta\left(p\right)\varphi_\alpha\left(p\right)\ dp}\\=
\exp{\left(-\left|\alpha-\beta\right|^2 /2 +
\left(\alpha^*\beta-\alpha\beta^*\right) /2
\right)}.
\end{array}
\]
Wave functions of field modes
\[
\begin{array}{l}
\phi^{\left(\pm\right)}_{\alpha,n}\left(x,t\right)\\
=\int{u^{\left(\pm\right)}_p }\varphi_{\alpha,n}\left(p\right)\exp{\left(ipx\mp i\varepsilon\left(p\right)t\right)}\ dp
\end{array}
\]
give the set of representations of quantum field associated to each set of wave functions

\begin{equation}
\begin{array}{l}
	\hat{\psi}\left(x,t\right)=\\
	\sum_{n=-\infty}^{\infty}{
	\phi^{\left(+\right)}_{\alpha,n}\left(x,t\right)\hat{a}_{\alpha,n}+
	\phi^{\left(-\right)}_{\alpha,n}\left(x,t\right)\hat{b}^+_{\alpha,n}
	}
	\end{array}
\end{equation}

Communication relations are canonical in each representation 
\[\left\{\hat{a}_{\alpha,n}\hat{a}_{\alpha,m}^+\right\}
=\left\{\hat{b}_{\alpha,n}\hat{b}_{\alpha,m}^+\right\}=\delta_{n,m}
\]
and non-canonical for operators attributed to different points of phase space.
For example
\[\left\{\hat{a}_{\alpha,0}\hat{a}_{\beta,0}^+\right\}
=\exp{\left(-\left|\alpha-\beta\right|^2 /2 +
\left(\alpha^*\beta-\alpha\beta^*\right) /2
\right)}
\]

All representations have common vacuum state $\hat{a}_{m}\cat{0}=0$ and $\hat{b}_{n}\cat{0}=0$ for all $n$. 
Transforms between representations do not entangle the creation and annihilation operators, thus the vacuum state is common for all the representations -- it is not needed here to distinguish between the representations, and there is no vacuum noise.

\begin{figure}[htp]
\setcaptionmargin{15mm}
\onelinecaptionsfalse 
\includegraphics[width=1.8in]{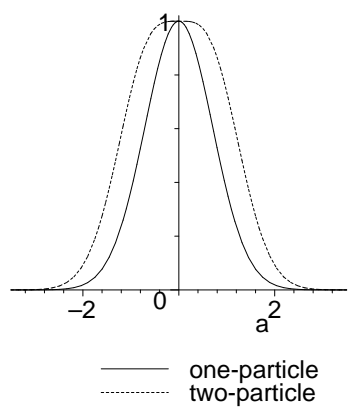}
\captionstyle{normal}\caption{Probability of particle registration versus distance from position of state. Two-particle state looks more extensive then the one-particle one. }
\label{dist}
\end{figure}

One-particle state $\cat{0,1_0}=\hat{a}^+_{0,0}\cat{0}$ in the origin $\alpha=0$ is not one-partical state out of the origin
\[P_{\beta,0}\left(1\right)=\brc{0,1_0}{\hat{a}^+_{\beta,0}\hat{a}_{\beta,0}}
=e^{-\left|\beta\right|^2},
\] 
and for two-particle state $\cat{0,1_0,1_1}=\hat{a}^+_{0,0}\hat{a}^+_{0,1}\cat{0}$  
probability of registration
\[P_{\beta,0}\left(1\right)=\left(1+\left|\beta\right|^2\right)e^{-\left|\beta\right|^2},
\]
 depends on distance from the origin as it is shown in graph (\ref{dist}).

\begin{figure}[ht]
\setcaptionmargin{0mm}
\onelinecaptionsfalse 
\includegraphics[width=1.8in]{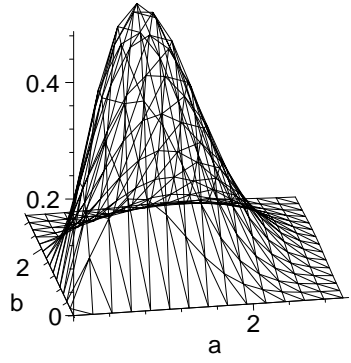}
\captionstyle{normal}\caption{Correlation function of joint registration of two particles for two-particle state. It depends from distances $a$ and $b$ between detectors and state origin.}
\label{gcor}
\end{figure}

Probability of joint registration of two particles is characterized by correlation function 
\begin{equation}
\begin{array}{r}
	C\left(\alpha,\beta\right)=
\brc{1_1,1_0}{\hat{a}^+_{\beta,0}\hat{a}_{\beta,0}\hat{a}^+_{\alpha,0}\hat{a}_{\alpha,0}}\\
-\brc{1_1,1_0}{\hat{a}^+_{\beta,0}\hat{a}_{\beta,0}}\\
\cdot\brc{1_1,1_0}{\hat{a}^+_{\alpha,0}\hat{a}_{\alpha,0}}
	\end{array}
\end{equation}

shown in the following graph (\ref{gcor}). 

The reason of correlation at registration of particles in given state is in not enough correspondence of detectors simulated by particle number operators to the state. In the case of complete correspondence (the origin) there is no correlation.

\section{Summary}
Imaginarity of entanglement can be caused by one of the next reasons:
\begin{itemize}

\item Standard scheme of probability definition leads to the imaginarity of entanglement because of restriction of wave packets;

\item Incomplete localization leads to imaginarity of entanglement because of the not enough correspondence between states and detectors.
\end{itemize}


\nopagebreak

\end{document}